\documentclass{svjour2}                    
\smartqed  
\usepackage{graphicx}
\newcommand{\beqn}{\begin{eqnarray}}
\newcommand{\eeqn}{\end{eqnarray}}

\journalname{Foundations of Physics}
\begin{document}

\title{Testing super-deterministic hidden variables theories}


\author{Sabine Hossenfelder }


\institute{S. Hossenfelder \at
              NORDITA, Roslagstullsbacken 23, 106 91 Stockholm, Sweden \\
              \email{hossi@nordita.org} }

\date{Received: date / Accepted: date}

\maketitle

\begin{abstract}
We propose to experimentally test non-deterministic time evolution in quantum
mechanics by consecutive measurements of non-commuting observables
on the same prepared state. While in the standard theory the measurement
outcomes are uncorrelated, in a
super-deterministic hidden variables theory the measurements would
be correlated. We estimate that for macroscopic experiments the
correlation time is too short to have been noticed yet, but that
it may be possible with a suitably designed microscopic experiment 
to reach a parameter range where one would expect a super-deterministic
modification of quantum mechanics to become relevant. 

\end{abstract}

\section{Introduction}

Quantum mechanics in the standard Copenhagen 
interpretation is fundamentally non-deterministic. The collapse of the wave function 
cannot be predicted, and outcomes 
of measurements are merely probabilistic. Because of its implications for 
the nature of true coincidence this issue has caused plenty discussions. The 
free will, even consciousness itself, has been placed within this 
randomness, for within a truly deterministic world the future might be unknown 
to us, but is in principle entirely determined by the past and unchangeable through 
human action -- it is the world of Laplace's demon. 

Despite the philosophical and interpretational upheaval, physicists 
have come to live with the quantum mechanical non-determinism in a very pragmatic style. 
It is without doubt that quantum mechanics and its younger cousin quantum 
field theory are experimentally extremely well confirmed. But the aim of our 
scientific endeavors is pushing beyond what we already know, and the axioms of quantum mechanics 
left us with the question whether this non-determinism is really fundamental, 
or whether it may arise from an underlying deterministic theory. If it was,
our inability to predict the outcome of a measurement could be due to an incomplete 
knowledge of the initial state, the complete knowledge existent but unavailable 
to us, encoded in what is commonly named `hidden variables.'

For a locally causal theory, an explanation of quantum effects through 
hidden variables can be shown to be in disagreement with experiment 
via Bell's theorem \cite{Bell:1964kc}, or its generalization respectively \cite{Clauser:1969ny}. 
Many tests \cite{Aspect:1982fx,Weihs:1998gy,Tittel:1998ja,Rowe,Moehring,Ansmann,Hasegawa} have meanwhile 
been performed and found that the hidden variables theories
for which Bell's theorem applies are not realized in nature. There are various loopholes
in the conclusions that can be drawn from the set-up of these experiments and not all of these
loopholes have yet been satisfactory closed (for a recent summary see e.g. \cite{loop}). But even so,
local hidden variables are strongly disfavored. However, Bell's theorem relies 
on the assumption that one has the freedom to choose the detector's settings without modifying the 
prepared state that one wishes to measure. In case this freedom is not given, the
conclusion of Bell's theorem does not apply. Theories that violate this assumption have become known 
as `super-deterministic,' and usually do not find much of an appreciation 
because they seem to rely on a `conspiracy' that is an inability of the
experimentalist to do as he wishes with his detector. 

However, despite our unease with super-determinism it might be 
the way nature works. We will consider here such super-determinism
from a very general point of view and propose a possibility to experimentally test and constrain
such theories. This is an investigation of interest not only with regard to the foundations 
of quantum mechanics, but also for our ongoing search for a theory of quantum
gravity: Possibly our inability to unify quantum theory with classical
general relativity is due to our incomplete understanding of quantum mechanics.

\section{Super-determinism}

Let us start with clarifying which type of theory we will be examining. It is
not our aim to construct a detailed model here. Instead, we want to test 
super-deterministic hidden variable theories ({\sc SDHVT}) in a way as model-independent as possible, much like Bell's theorem
tests for hidden variables by only making a few assumptions rather than using
a specific model.

We will consider our space-time as a differentiable 4-manifold with a
3+1 slicing into space-like hypersurfaces $\Sigma_t$. An initial state
$\psi(t,x)$ on $\Sigma_t$ can be evolved forward and backward by help
of the evolution equation, or the Hamiltonian respectively, which is
an axiom of the theory. This evolution
is deterministic also in quantum mechanics. The non-deterministic ingredient of
the Copenhagen interpretation
is the probabilistic computation of measurement outcomes from the initial 
state, given by Born's rule: The outcome of a measurement of an
observable described by an operator $\hat O$ is an eigenvalue $O$ of that
operator $\hat O \psi_O = O \psi_O$, where $\psi_O$ is the corresponding
eigenvector. The probability $P(\psi, O)$ of measuring a 
state $\Psi(t,x)$ at value $O$ is then given by the square of the projection of the
state on the eigenvector
\beqn
P(\psi, O) = \left| \langle \psi | \psi_O \rangle \right|^2 \quad,
\eeqn
and the expectation value of the measurement is
\beqn
E(\psi, O) = \langle \psi \left| \hat O \right| \psi \rangle \quad.
\eeqn

In a {\sc SDHVT}, the initial state depends 
on additional parameters $\lambda_i$, and the outcome of the measurement is a function of the
hidden variables $\lambda_i$. If all $\lambda_i$ were known, the outcome
of the measurement could be predicted in principle. We will assume that there are finitely
many such hidden variables, $i \in \{1\dots N\}$. To good precision such a {\sc SDHVT} has to
reproduce Born's rule when the hidden variables are unknown and are
considered to be stochastically distributed over typical values describing the setting. 

The wave-function contains both the state that one wants
to measure and the detector on the slice $\Sigma_t$ -- in fact a separation
might not be uniquely possible. Normally both parts of the wave-function
are considered as separate entities. Super-determinism comes
in through the impossibility to change the part of the wave-function 
describing the detector independently from the one describing the prepared
state. This violates a central assumption of Bell's theorem. 

A simple example 
to understand such a feature is to 
consider a wave-function $\psi(t, x, \lambda_i)$ that is analytic on the whole slice $\Sigma_t$. It is
then not possible to prepare any wave-function on compact support, and to add non-overlapping
wave-functions. Instead, a different
setting of the detector necessarily implies a different prepared state, even though
both are spatially separated.
This change of initial conditions in the detector affecting the initial
conditions of the prepared state might be small, but it affects 
the hidden variables. 

One can now see why such super-deterministic theories are often dubbed `conspiracy theories.' 
In a certain sense, the detector already `knows' it will be detecting the
state that is being prepared. Not because there is backward causation
though -- causality is maintained in
its usual form -- but because it is impossible to prepare the state independently
of the detector to begin with. 
If this dependence goes unnoticed, it results in a seemingly non-deterministic 
measurement outcome. The impossibility to prepare state and detector independently
is not specifically due to an inability of the conscious beings conducting the 
experiments, but it is a consequence of fundamental determinism. 

At first sight, such super-determinism might seem nonlocal. 
To clarify the notion of local causality, let us examine whether it is locally causal according 
to Bell's definition:
\begin{quote}
 {\it ``A theory will be said to be locally causal if the probabilities attached to values of local beables in a space-time region 1 are unaltered by specification of values of local beables in a space-like separated region 2, when what happens in the backward light cone of 1 is already sufficiently specified, for example by a full specification of local beables in a space-time region 3...''} \cite{Bell} p. 239-40
\end{quote}

\begin{figure} 
\hspace*{1.0cm}
\includegraphics[width=6cm]{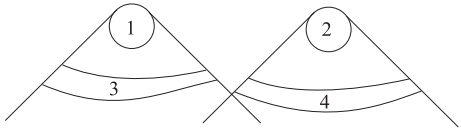}
 
\caption{Local causality \label{fig1}}
\end{figure}

(See Figure \ref{fig1}, also \cite{lc}). Thinking of our example with the analytic function, super-determinism can thus be locally causal 
according to the second part of the definition ({\it ``what happens in the backward light cone of 1 is already sufficiently specified, for example by a full specification of local beables in a space-time region 3''}) but not according to the first part ({\it ``the probabilities attached to values of local beables in a space-time region 1 are unaltered by specification of values of local beables in a space-like separated region 2.''}) The violation of
the second part of the definition is exactly what makes it impossible to chose the detector settings without altering the state that one wants to observe. It does however a priori not 
necessitate superluminal exchange of information or action at a distance.


In the following we will assume that the hidden variables in our theory are 
environmentally induced by which we mean their origin is in the impossibility
to treat the prepared state as independent from the rest of the experiment.
The environment is thus constituted of what is necessary to completely
specify the experiment, including the detector and any apparatus
necessary to prepare the state and to set up the experiment. It does not
include further non-necessary additions, such as for example an actual
observer reading out the measurement outcome. This is a minimalistic
assumption about the origin of the hidden variables, and a crucial one to
specify the sort of theory we are interested in on which
we will comment in the discussion. Note that this notion
of environment becomes ambiguous in ordinary quantum mechanics. There, it
would be possible to not only change the settings of the detector
as one wishes, but it would in principle be possible to change the whole
detector after one had set up the experiment. Then one would be faced
with the question if one had to consider all possible detectors to be
part of the possible environment. Luckily, in the {\sc SDHVT}, we do not
have this complication, since the detector cannot
be altered independently of the prepared state without changing
the whole experiment. 

\section{Testing Super-determinism}

We can now make some general considerations about these super-deterministic hidden variables
theories. Based on this, we can then propose experimental tests. 

The decisive feature of a deterministic hidden variable theory is that
knowing the variables allows one to predict the measurement outcome. To investigate
this feature, what we will
be interested in here is the correlation between measurements on identically
prepared states for which also the hidden variables are identical. In ordinary quantum mechanics, 
the outcome of measurements on 
identically prepared states will be statistically distributed according to Born's rule,
and entirely uncorrelated. In a hidden variables
theory, the outcomes of these measurements may be unknown if the hidden variables are unknown,
but the measurement outcomes will be the same. 

The difference between the both cases can be illustrated with a simple example. Consider
you are pregnant with twins (`identically prepared') and are given a risk estimate of $r$ 
(probability of measurement), based on your and your partner's family 
history, for a genetic disease of your offspring. If the twins are fraternal (corresponding to usual quantum mechanics), the risk 
that both children will have the disease is $r^2$. In this analogy,
quantum mechanics would imply there is no way to make states more `identical' than having 
the same parents -- there are no hidden variables. If the twins are identical (corresponding to the hidden
variables theory) the risk that both twins have the disease is $r$. On the risk of overstretching this
example, if you were given data of a large group of women with either all identical or all fraternal twins, you could
find out which case you are dealing with by studying the correlation of diseases among the twins. 
That, in essence, it what we will do.

Let us denote with $\kappa \in \{1...n\}$ a sequence of $n$ identically prepared states $\phi_\kappa$ and $P(\phi_\kappa, O)$
the probability to measure value $O$ for observable $\hat O$. For simplicity we assume that
the expectation value  $E(\phi_\kappa,O) =0$. Then, in ordinary
quantum mechanics the joint probability for two subsequent measurements of the same value $O$ is
\beqn
P(\phi_\kappa, O \wedge \phi_{\kappa +1}, O) =  P(\phi_\kappa, O)^2 \quad, \label{qm}
\eeqn
whereas in the {\sc SDHVT}
\beqn
P(\phi_\kappa, O \wedge \phi_{\kappa +1}, O) = P(\phi_\kappa, O) \quad. \label{sd}
\eeqn
The state $\phi_\kappa$ carries all the available information about the system,
i.e. in Eq. (\ref{qm}) it is the usual wave-function, whereas in Eq. (\ref{sd}) it is a
function also of the hidden variables. The above means in other words that the correlation
\beqn
{\mathrm{Corr}}(\nu,\kappa) = \frac{E(\phi_{\nu}, O \wedge \phi_{\kappa}, O  )}{E(\phi_\nu, O^2)}
\eeqn
does in the {\sc SDHVT} not vanish for $\nu \neq \kappa$, whereas in standard quantum mechanics
it does. We will denote with ${\mathrm{Corr}}_\kappa = {\mathrm{Corr}}(0,\kappa)$ the
comparison to the first state in the sequence.

The problem with detecting this modification of quantum mechanics 
is that to produce an identically prepared initial state in the hidden variables theory, 
one would have to make sure the hidden variables have identical values too, which, 
given that they are hidden, is difficult to achieve. In a non-ideal setup, which
one generally will be faced with, the correlation ${\mathrm{Corr}}_\kappa$ lies between 0 and 1,
tending towards the usual quantum mechanical value of zero the less reliably one can
expect the states to be identical. The important point is that this correlation time is
measureable, and any value larger than zero is in contradiction with standard
quantum mechanics.

The number of hidden variables, $N$, that describe the setting of prepared state and detector
will increase the more degrees of freedom the system under consideration has which means
in most cases, the larger it is. This is the reason
why it is plausible to assume the hidden variables are statistically distributed,
resulting in a seemingly random measurement outcome. The more variables we have to
take into account, the more difficult it will be to measure any correlation. 

The
hidden variables come in through both the preparation of the state and the detector,
from which the preparation of the state is the more problematic part because it
usually involves an even larger system. Luckily, this latter problem can be circumvented
by doing repeated measurements on the same state. This requires to chose a setting
in which the state (at least with some probability) is returned into the initial state.
In this case we are then interested in the correlation of measurements of the
same observable $\hat O$ of the same state $\phi$ at time steps labeled by $\kappa$.
With a slight abuse of notation, we are thus interested in the autocorrelation of
the time series:
\beqn
{\mathrm{Corr}}_\kappa = \frac{E(\phi, O, t=0 \wedge \phi, O, t=\kappa )}{E(\phi, O^2)}~.
\eeqn
It then remains the task of
choosing repeatable measurements in between which the detector itself undergoes as few 
changes as possible. 

For the autocorrelation we make the usual ansatz of exponential decay
\beqn
{\mathrm{Corr}}_\kappa = \exp \left( - \kappa/\tau \right) ~,
\eeqn
where $\tau$ is some timescale, the autocorrelation time, encoding the rate of change of the experimental
setup and thus the environmentally induced hidden variables. For this ansazt we have assumed that the
divergence of the system from a perfect repetition happens by incremental disturbances that are described 
by a homogeneous Poisson-process, i.e. the changes that lead to the decay of the autocorrelation 
are statistically independent and follow a probability distribution
which is uniform in time. The larger the setting and the shorter the typical timescales on which its degrees of freedom change, 
the smaller $\tau$ and the faster the correlation will decrease. In the limit $\tau \to 0$,
one reproduces standard quantum mechanics. 

With these considerations, we can now identify the following criteria as useful for the 
purpose of constraining {\sc SDHVT}s:

\begin{enumerate}
\item Instead of measuring a sequence of individually prepared states, chose a setting
in which the state (at least with some probability) is returned into the initial state
and repeated measurements on the same state can be performed. 
\item The experimental setup itself and the detector should be as small as
possible to minimize the number of hidden variables (i.e. $N$ should be small). 
\item The repetition of measurements should be as fast as possible so any
changes to the hidden variables of the detector in between measurements are minimized (i.e. $\kappa < \tau$).
\end{enumerate}

\section{Proposed Experiment}

The quantity ${\mathrm{Corr}}_\kappa$ will be a function of those hidden 
variables that change during the time interval $\kappa$. Since we assumed
these variables to be induced by the environment that, in the super-deterministic
theory, can no longer be treated as independent from the prepared state,
the correlation time will depend on the experimental setup. We will here
propose a general outline for an experiment that should be doable with
technology available today, and allow a measurement that constrains 
the {\sc SDHVT}.

\begin{figure} 
\vspace*{-0.0cm}
\includegraphics[width=7cm]{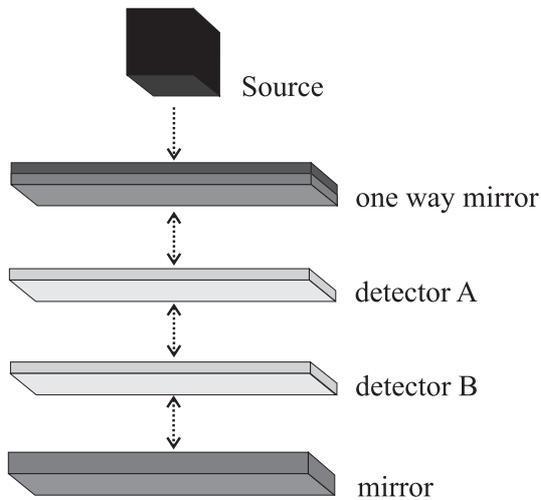}
 
\caption{Schematic setup of experiment.
\label{fig2}}
\end{figure}

The setup of a possible experiment is shown in Figure \ref{fig2}. First, one
needs a source for single particles, most conveniently electrons or photons. These
particles enter through a one-way mirror a system consisting of two 
detectors, A and B, behind which there is another mirror. The particle
then, ideally, enters a loop. The two detectors 
measure two non-commuting variables, $A$ and $B$, with $[A,B] \neq 0$, in 
a way that, according to standard quantum mechanics, measurement of one variable should 
destroy information obtained in the previous
measurement of the other variable. Since actual one-way mirrors do not exist, one should understand this as 
a material with a small chance of the particle trespassing, say $p=1\%$. Then, only
one out of 100 particles will on the average enter the system, but once it has
done so, it has a 99\% probability of staying for the next loop.


To see whether such an experiment is possible in an interesting parameter range, 
let us estimate what is the typical decay time to 
expect.
As an example, consider a photodetector constituted of $N$ atoms that measure an infalling
photon by excitation of an electron into the conduction band. The energy gap of the 
material be $\Delta E$ and the temperature be $T$.
At that temperature, one single atom has a probability of $\exp(-\Delta E/T)$
to become excited by thermal motion, and it remains so for the average 
electron-hole recombination time $\tau_{\mathrm r}$. The 
time $\tilde \tau$ for $N$ atoms to undergo a statistical change is thus
\beqn
\tilde \tau \approx \frac{\exp(\Delta E/T)~ }{N} \tau_{\mathrm r} ~.
\eeqn
There are many other sorts of noise, but this is the one with the highest
frequency, and thus the one relevant for the decay time.

We set this time in relation to the decay time of the correlation by 
$\tau = \alpha \tilde \tau$. Here, the dimensionless parameter $\alpha$ is the parameter 
we want to constrain by experiment. For a {\sc SDHVT} one would expect it to
be of order one, whereas for ordinary quantum mechanics it is zero.

If one inserts typical values of $\Delta E \sim .5$eV, $T \sim 300$K, $\tau_{\mathrm r} \sim 1$~ns,
and $N \approx 10^{20}$ one finds $\tilde \tau \sim 10^{-20}$s and,
for $\alpha \approx 1$, a hopelessly
small correlation time. One might think about cooling the whole
system down. However, this would practically necessitate immersion in some
liquid, thereby increasing $N$ and the difficulty to perform an experiment
to begin with. However, if we consider instead a microscopically small detector,
maybe with an extension of some $\mu$m, we could get down to 
$N \approx 10^{15}$. Let us further take
a semi-conductor with a fairly large band gap of $\Delta E \approx 1$~eV.
Then we get down to $\tilde \tau \approx 10^{-6}$s. This now has to be
compared with the typical time in which measurements can be repeated. 

In $10^{-6}$sec a photon travels a distance of about 100m. The
experiment cannot be arbitrarily small because it still should
be larger than the
wavelength of the photon, but a size of a mm is, with visible light, still large
enough. In this case, the time for a photon to get from one end of
the experiment to the other is below $10^{-11}$sec, and 
is as desired much shorter than the decay time, leaving space for
non-ideal material and other sources of experimental uncertainty. 

For the example of the 
photon, the detectors A and B could measure different polarization states in a photonic 
version of the Stern-Gerlach experiment, when such devices can
be made sufficiently small. Alternatively, one can use electrons and
spin-filters \cite{spinfilter}. While 
the research in this area is not yet sufficiently advanced, it may
soon be. To return the electron into
the initial state, the function of a mirror could
for example be mimicked by Bragg-reflection on sufficient layers of suitably 
spaced atoms. 
Electrons bring the advantage that the
wavelength is smaller and thus the experiment can be made smaller. It has
the disadvantage that it is more difficult to handle charged particles. 

In either case, the detector would let the particle (photon or electron)
trespass straight for one particular combination of measurement outcomes 
(polarization or spin), while the particle would be deviated to
leave the loop if it does not have this particular combination. The quantity
of interest is then time it takes for the particle to leave the loop at
either detector A or B. The average value of these time measurements is 
the correlation time.

Let us
denote the measurement outcomes the particle needs to have to stay
in the loop with $a$ and $b$, corresponding to the two detectors (for
example spin up and left).
The probabilities for these measurement outcomes be $p_a$ and $p_b$,
respectively.
In ordinary quantum mechanics the probability for a particle that 
has entered the system to be deviated out of the system at detector
A or B before it
returns to the point of entry is $(1-p_a)^2 (1-p_b)$. Then, it has a 
probability of $p$ that it leaves through the imperfect one-way mirror, and 
the probability to make it into the next loop is $(1-p)(1-p_a)^2(1-p_b)$. 
The probability that the particle is still in the system after the $m$th 
reflection at the one-way mirror is $(1-p)^{m} (1-p_a)^{m+1} (1-p_b)^m$. 
In the {\sc SDHVT} however, we know that if the particle has fulfilled the
conditions to make the loop once it will continue to do so and, on a duration shorter than
the autocorrelation-time, it should return to the mirror on the
second screen with probability $(1-p)^m (1-p_a) (1-p_b)$. 

With $p\ll 1$ and $p_a, p_b$ of order one, in ordinary quantum mechanics the probability
is thus very high that the photon will be detected within only a few rounds. In 
the {\sc SDHVT}, detection should occur either in the first round,
or with a delay relative to the standard case. There is not much
use in making the propagation time of the particle very short by
making the experiment very small, since the time still has to be
measureable. However, with present-day technology, it seems feasible
to measure a correlation time of $10^{-6}$sec and distinguish it
from no correlation.

In addition to the above considerations about measurability of the autocorrelation,
there is of course experimental error that needs to be taken into account. Most
crucially, we have so far assumed that the mirrors are perfectly even. In reality, they
will of course have a finite surface roughness and the electron or photon will only return
to its initial state with some limited precision, which then alters the 
probability for it to make the same loop again. We can estimate this effect by
noting it will become relevant if the unwanted dislocation of the particle per loop, 
$\Delta z_m$,  
through a non-perfect reflection makes a non-negligible change to its relative 
location to the degrees of freedom of the environment. In the case considered
here, the relevant distance would be of about one atomic diameter, or $10^{-10}$m.
If we denote this distance with $z_N$, then we have an additional contribution
to the autocorrelation time that increases with $m$,
\beqn
Corr_{\kappa} = \exp \left( - \frac{\kappa}{\alpha \tilde \tau} \right) + \exp \left(- m \frac{\Delta z_m}{z_N} \right) \quad.
\eeqn

\section{Discussion}

The here proposed test aims directly at falsifying the non-deterministic nature
of quantum mechanics. Since we know that local theories with hidden variables 
are strongly disfavored by experiment already, the type of model that could replace
standard quantum mechanics considered here is non-local in
the specific sense of being super-deterministic, i.e. the prepared state
is correlated with the detector. 

Of course any proposed test is only as good as the assumptions going into
deriving the constraints, so let us summarize them. Most crucially, 
we have made the minimalist assumption that the hidden
variables stem from the correlation with the detector and possibly other
parts of the experimental setup. This is a restriction to a subclass of
{\sc SDHVT}s since in principle the remainder of the universe might contain additional variables relevant
to the evolution of the subsystem we are considering. While this is possible,
this would be a type of {\sc SDHVT} that is in practice not testable anyway, 
so we have focused here on a version of {\sc SDHVT} that, if realized in nature, 
could falsify non-determinism in quantum measurement. (This is not to say
that the type of theory considered in this paper is the only type of {\sc SDHVT} 
which is testable.) We also have not addressed here the possibility of a
combination of a super-deterministic evolution with a limited version of free will,
a possibility that merits mentioning  since it has recently been shown that 
allowing for a small initial correlation between measurement device and 
measured system is sufficient to reproduce nonlocal correlations \cite{partial}.

Further, we have implicitly made the assumption that variables that are not
relevant for the description of the experiment in the standard theory
are not among the hidden variables either. This includes for example that
the space-time the prepared state and the detector are embedded in does 
not carry additional variables which would have to be taken into account for 
the measurement outcome, and neither does the description of the detector's
constituents on scales smaller than relevant for the experiment add to the
hidden variables. Another assumption we have made is that we explicitly
used there are finitely many hidden variables. We also used an ansatz
according to which the correlation decays exponentially in time. 

It should also be noted that the assumption of a super-deterministic
correlation between prepared state and detector is not in conflict with
a local time evolution assuming that their past lightcones intersect. However, 
this assumption will always be fulfilled for Earth based experiments.

Finally, it should be noted that we have assumed that the commutators
between the measured observables remain the same as in standard quantum
mechanics.

\section{Closing Remark}

During the work on this manuscript, it was brought to my attention 
footnote 1 of E.~P.~Wigner's paper \cite{Wigner:1976ga}:
\begin{quote}
{\it ``Von Neumann often discussed the measurement of the spin component
of a spin-$\frac12$ particle in various directions. Clearly,
the possibilities for the two possible outcomes
of a single such measurement can be easily accounted for
by hidden variables [...] However, Von Neumann felt that this is
not the case for many consecutive measurements of the spin component
in various different directions. The outcome of the first such
measurement restricts the range of values which the hidden parameters
must have had before that first measurement was undertaken. The
restriction will be present also after the measurement so that the
probability distribution of the hidden variables characterizing the spin will
be different for particles for which the measurement gave a positive
result from that of the particles for which the measurement gave a
negative result. The range of the hidden variables will be
further restricted in the particles for which a second measurement
of the spin component, in a different direction, also gave a
positive result. A great number of consecutive measurements will
select particles the hidden variables of which are all so
closely alike that the spin component has, with a high probability,
a definite sign in all directions. However, according to quantum
mechanical theory, no such state is possible. 

Schr\"odinger
raised the objection against this argument that the measurement
of a spin component in one direction, while possibly specifying
some hidden variables, may restore a random distribution of some other
hidden variables. It is this writer's impression that Von Neumann did not
accept Schr\"odinger's objection. His point was that the objection
presupposed hidden variables in the apparatus used for the 
measurement. Von Neumann's argument needs to assume only two
apparata, with perpendicular magnetic fields, and a succession of
measurements alternating between the two apparata. Eventually, even
the hidden variables of both apparata will be fixed by the outcomes
of many subsequent measurements of the spin component in their respective
directions so that the whole system's hidden variables will be fixed.

Von Neumann did not publish this apparent refutation of Schr\"odinger's
objection.''}
\end{quote}
In a later volume of the same journal J.~F.~Clauser commented on
Wigner's recollection of Von Neumann's and Schr\"odinger's discussion
and provided a particular example for hidden variables which 
are able to reproduce the predictions of
quantum mechanics for repeated measurements of polarization directions \cite{Clauser}. In
his reply to Clauser, Wigner (ibid) pointed out that Clauser's example
comes at the cost sacrificing two-way causality, i.e. there is no one-to-one
correspondence between states at two instants of time and no
time-reversal invariance. To this Clauser replied (ibid) that time-reversal
invariance is maintained if one takes into account the history of
the measurements. Clausers example, though admittedly contrived and
applicable only for a special case, thus represents
a case of hidden variables not of the type discussed here, one in which
the hidden variables have nothing to do with the correlation between
prepared state and detector, and the correlation time is zero.

The here proposed test would be a realization of Von Neumann's
thought experiment, taking into account that the alternating
measurements can in practice not be done on exactly the same
state. Thus, instead of entirely fixing the outcome of measurements,
the effect of the hidden variables is a non-vaninishing 
correlation time between measurement outcomes.

\begin{acknowledgements}

I thank Joy Christian and Stefan Scherer for helpful comments, and Chris Fuchs for feedback as well as
for pointing me towards reference \cite{Wigner:1976ga}. 
\end{acknowledgements}

\end{document}